\documentclass[10pt]{article}
\usepackage{fullpage}
\usepackage{amsmath}
\usepackage{amsfonts}
\usepackage{amssymb}
\usepackage[dvips]{epsfig}
\usepackage{color}

\def\blue{\textcolor{black}}

\def\##1{\underline{#1}}
\def\=#1{\underline{\underline{#1}}}

\def\+#1{\underline{\bf #1}}
\def\*#1{\underline{\underline{\bf #1}}}

\def\r#1{(\ref{#1})}
\def\l#1{\label{#1}}
\def\c#1{\cite{#1}}

\def\le{\left(}
\def\ri{\right)}
\def\les{\left[}
\def\ris{\right]}

\def\.{\cdot}

\def\epso{\epsilon_{\scriptscriptstyle 0}}

\def\muo{\mu_{\scriptscriptstyle 0}}

\def\eps{\epsilon}

\begin{document}

\begin{center}

\LARGE{ {\bf Towards a piecewise--homogeneous  metamaterial model of the collision of two  linearly polarized gravitational
plane waves}}
\end{center}
\begin{center}
\vspace{10mm} \large

 Tom G. Mackay\footnote{E--mail: T.Mackay@ed.ac.uk.}\\
{\em School of Mathematics and
   Maxwell Institute for Mathematical Sciences\\
University of Edinburgh, Edinburgh EH9 3JZ, UK}\\
and\\
 {\em NanoMM~---~Nanoengineered Metamaterials Group\\ Department of Engineering Science and Mechanics\\
Pennsylvania State University, University Park, PA 16802--6812,
USA}\\
 \vspace{3mm}
 Akhlesh  Lakhtakia\footnote{E--mail: akhlesh@psu.edu}\\
 {\em NanoMM~---~Nanoengineered Metamaterials Group\\ Department of Engineering Science and Mechanics\\
Pennsylvania State University, University Park, PA 16802--6812, USA}

\normalsize

\end{center}

\begin{center}
\vspace{15mm} {\bf Abstract}

\end{center}
We considered the experimental realization of a Tamm medium that is optically equivalent
to the collision of two  linearly polarized gravitational
plane waves as a piecewise-homogeneous metamaterial. Our formulation
 was based on the homogenization of remarkably  simple arrangements of oriented ellipsoidal nanoparticles
 of isotropic dielectric--magnetic mediums.
The inverse Bruggeman homogenization was used to estimate the constitutive parameters, volume fractions, and shape parameters for the component mediums. The presented
formulation is appropriate for the regions of spacetime where the two gravitational  plane waves interact, excluding the immediate vicinity
of the
nonsingular Killing--Cauchy horizon
at the focusing point of the two plane waves.

\noindent {\bf Keywords:} colliding gravitational plane waves; homogenization; Killing--Cauchy horizon;
metamaterial; inverse Bruggeman  formalism

\section{Introduction}

Metamaterials are engineered materials which, through judicious design,
can facilitate the realization of exotic phenomenons such as
negative refraction and cloaking \c{Metamaterials}.
Metamaterials also present opportunities to study
 general-relativistic scenarios. This is because a
formal analogy  exists between light propagation in empty
curved spacetime and light propagation in a certain nonhomogeneous
anisotropic or bianisotropic medium, called a Tamm medium
\c{Skrotskii,Plebanski,SS}. For examples,
  theoretical descriptions of metamaterial
analogs of  black holes \c{Smolyaninov_NJP}, de Sitter spacetime
\c{Li_1,Li_2} including Schwarzschild-(anti-)de Sitter spacetime \c{ML_PRB},  strings \c{Li_3} including
 cosmic
strings \c{Spinning_string},  and wormholes \c{Greenleaf},   have been investigated. Such analogs may offer insights into spacetime scenarios which cannot be practicably studied by direct methods.

In theory, analogs of curved spacetime may be constructed from metamaterials. However,
achieving this  in practice presents a major  challenge to experimentalists.
Concrete experimental proposals are conspicuously absent from the literature, but there are some exceptions.
Lu \emph{et al.} \c{Lu_JAP} presented a detailed metamaterial representation
 of a two-dimensional black hole. Their metamaterial took the form of a homogenized composite material (HCM) comprising relatively simple component mediums.
We established a
metamaterial formulation for the Tamm medium representing
Schwarzschild-(anti-)de Sitter spacetime \c{ML_PRB}. In our approach the metamaterial was also a HCM, arising from the homogenization of isotropic dielectric and isotropic magnetic component mediums
which are distributed randomly as oriented
spheroidal particles.
In this paper, we apply the same approach to the case of
the Tamm medium corresponding to the  collision of two  linearly polarized gravitational
plane waves \c{Griffiths}.

\blue{The collision process is a complex one from the perspective of both theoretical and numerical analyses: its intrinsic nonlinearity
 gives rise to either  a spacetime singularity or a
 nonsingular Killing--Cauchy horizon at the focusing point of the two plane waves \c{Griffiths}.
As such, a convenient experimental analog could be particularly useful, as was succinctly pointed  by Bini \emph{et al.}  \c{Bini2014}.
Colliding gravitational plane waves may be used to study Cauchy horizons in classical general relativity \c{Ori,Yurtsever}, as well as
 string behaviour in gravitational fields \c{Vega_PRD,Jofre_PRD}. These plane waves may also assist in the study of black hole collisions \c{Ferrari_1988}
 and travelling waves on cosmic strings \c{Garfinkle_PRD}.}

\blue{
 Motivation for the present paper was
chiefly provided by a  recent study \c{Bini2014} in which the Tamm medium analogy was exploited to
 to highlight some interesting optical properties of the spacetime of two colliding gravitational plane waves.}

The plan of this paper is as follows: Section~\ref{Tamm-medium}
describes the Tamm medium for
the region of spacetime in which the two plane waves interact to form the nonsingular Killing--Cauchy horizon.
Section~\ref{iBr} succinctly describes the inverse Bruggeman homogenization formalism used for the piecewise-homogeneous
implementation of the Tamm medium, followed by illustrative numerical results in Sec.~\ref{nr}. The paper concludes with a discussion in Sec.~\ref{disc}. \blue{In the units adopted,  the Newtonian constant and the speed of light in vacuum are both equal to unity.}

\section{Tamm medium for colliding-gravitational-wave spacetime}\label{Tamm-medium}

The collision of two oppositely-directed gravitational planes waves  may be represented as an
exact solution of the
 Einstein field equations \c{Griffiths}. Following the collision, focussing effects give rise to
 either a nonsingular Killing--Cauchy horizon or a spacetime singularity.
  \blue{Conventionally, the corresponding spacetime, prior to the creation of the nonsingular Killing--Cauchy horizon or the spacetime singularity,   is partitioned into four regions: Region I,
 where the two plane waves interact; Regions II and III, each corresponding to a single plane wave before the interaction; and Region IV, which is a flat spacetime region representing the initial state before the passage of the two plane waves.}
We focus on Region I  bounded by
\begin{equation}
\left.
\begin{array}{l}
-t \leq z \leq t\\
\displaystyle{0 \leq t \leq \frac{\pi}{2}}
\end{array}
\right\},
\end{equation}
\blue{for propagation along the $z$ axis.}
 Here the instant of collision occurs at  $t=0$ while  at $t=\pi/2$ either a nonsingular Killing--Cauchy horizon or a spacetime singularity is created.

Provided that the plane waves are linearly polarized and they propagate in opposite directions along the $z$ axis,
the corresponding line element may be expressed as  \c{Ferrari_GRGa,Ferrari_GRGb,Ferrari_PRSLA}
\begin{equation} \l{Line_element}
ds^2 = {F}_+^2 (t) \le dt^2 - dz^2 \ri + \frac{{F}_- (t)}{{F}_+ (t)}\, dx^2 + \cos^2 (z) \, {F}_+^2 (t)\, dy^2,
\end{equation}
where the scalar function
\begin{equation}
{F}_\pm (t) = 1 \pm \sigma \sin \le t \ri.
\end{equation}
Herein $\sigma = \pm 1$, with $\sigma = +1$ corresponding to the nonsingular Killing--Cauchy horizon solution at  $t=\pi/2$ whereas
$\sigma = -1$ corresponds to the spacetime singularity solution at  $t=\pi/2$. In the following we consider the $\sigma = +1$ option; the prescription of a metamaterial analog for the $\sigma = -1$ case follows in a similar vein to the $\sigma = +1$ case but the constitutive-parameter regimes for  $\sigma = -1$  would be more challenging to realize in practice than those for  $\sigma = +1$.

The optical response of vacuum in the curved spacetime represented by the line element
\r{Line_element} is formally equivalent to the optical response of a spatially and temporally nonhomogeneous medium
 in flat spacetime \c{Skrotskii,Plebanski,SS}, known as the Tamm medium.
As recently shown by
Bini \emph{et al.} \c{Bini2014}, the Tamm medium here is an orthorhombic dielectric--magnetic medium \c{EAB}, which may be characterized by the  electromagnetic constitutive relations
\begin{equation}
\left.
\begin{array}{l}
\#D (x,y,z,t) = \epso \=\gamma(z,t) \cdot \#E (x,y,z,t)
\\
\#B (x,y,z,t)  = \muo \=\gamma (z,t) \cdot \#H (x,y,z,t)
\end{array}
\right\}, \l{CR}
\end{equation}
wherein SI units are implemented with $\epso = 8.854\times
10^{-12}$~F~m~$^{-1}$ and $\muo = 4\pi\times 10^{-12}$~H~m$^{-1}$, and
\begin{equation} \l{Tamm}
\=\gamma_{\,} (z,t) = \mbox{diag} \les \gamma^x_{} (z,t),\,
\gamma^y_{}(z,t), \,\gamma^z_{}(z,t) \, \ris\,,
\end{equation}
with
\begin{equation} \l{Tamm_e_m}
\left.
\begin{array}{l}
\gamma^x (z,t) = \displaystyle{  \cos \le z \ri \sqrt{\frac{{F}^3_+(t)}{{F}_- (t)}} } \vspace{8pt}\\
\gamma^y (z,t) = \displaystyle{ \frac{1}{\cos \le z \ri} \sqrt{\frac{{F}_- (t)}{{F}^3_+(t)}}} \vspace{8pt} \\
\gamma^z (z,t) =  \displaystyle{\cos \le z \ri \sqrt{\frac{{F}_- (t)}{{F}^3_+(t)}} }\end{array} \right\}.
\end{equation}

\begin{figure}[!ht]
\centering
\includegraphics[width=3.5in]{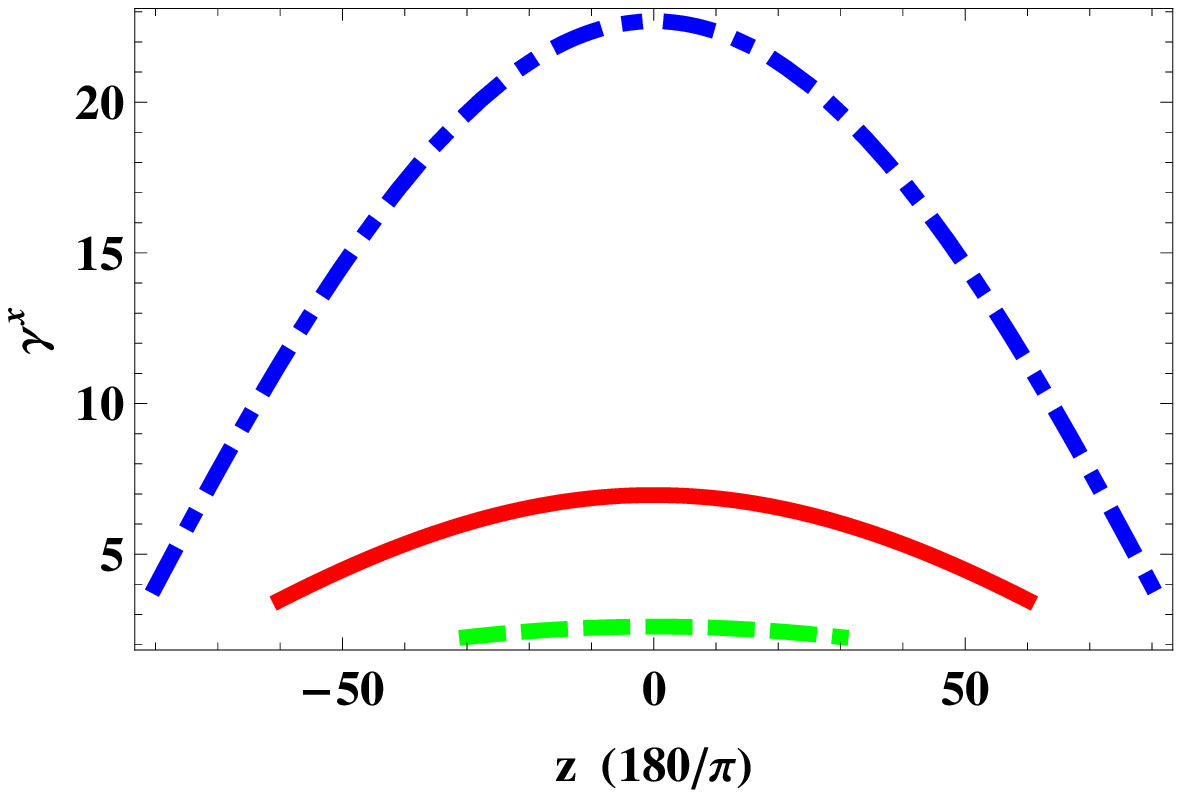}
\includegraphics[width=3.5in]{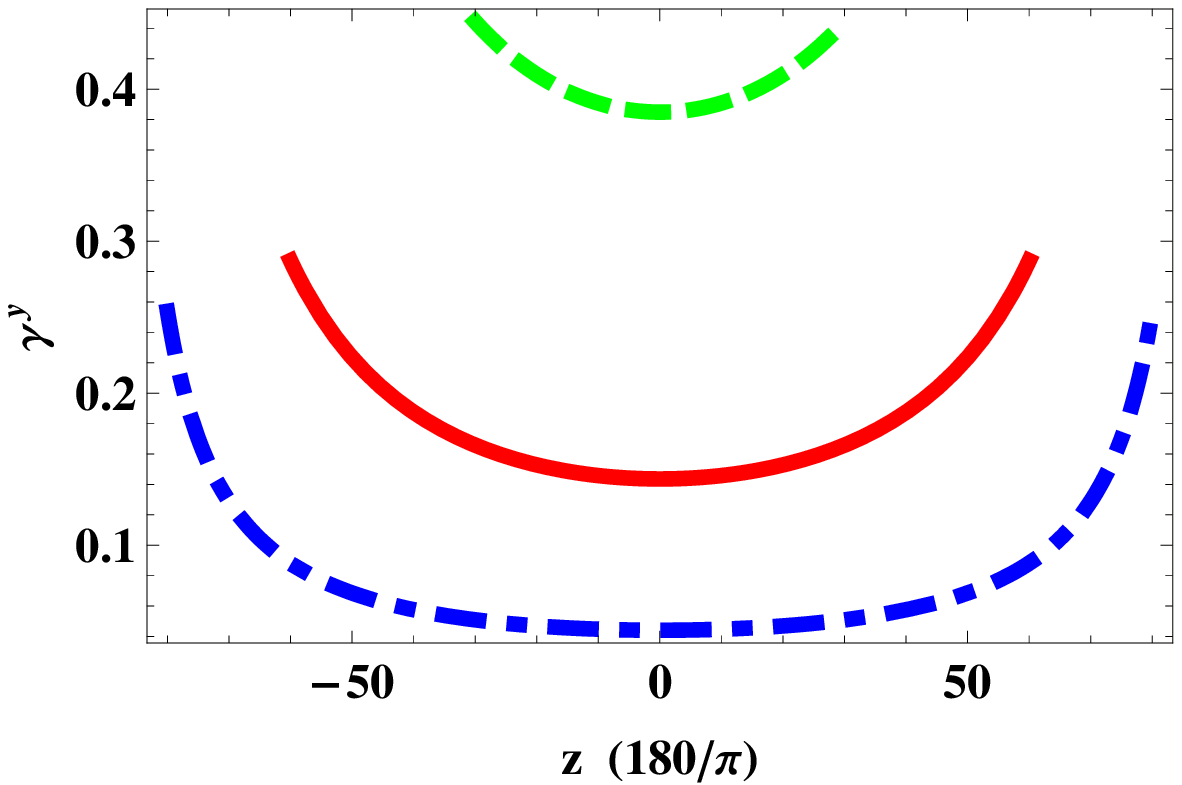}
\includegraphics[width=3.5in]{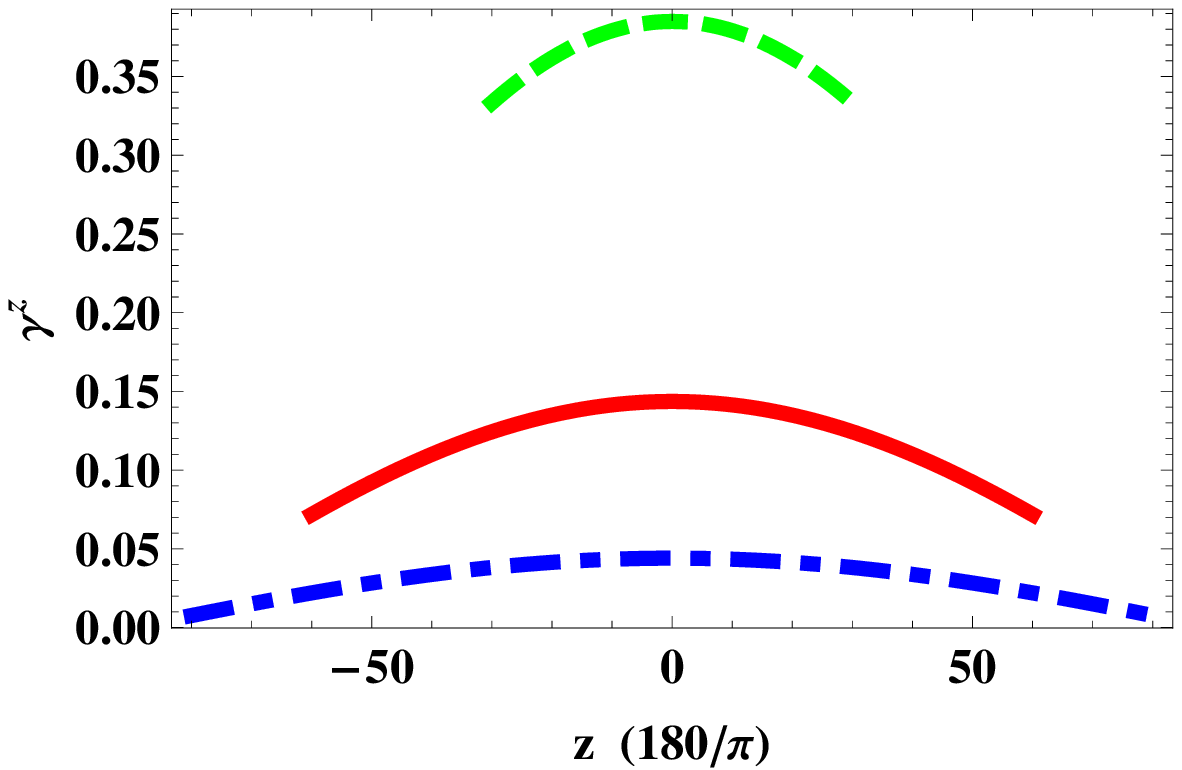}
 \caption{ The parameters $\gamma^{x,y,z}$  plotted versus $z \in \le -t, t  \ri$ for $t=  \pi/6$ (green dashed curves),
 $\pi/3$ (red solid curves), and $4 \pi /9$ (blue broken dashed curves).}
  \l{fig1}
\end{figure}

%

In Fig.~\ref{fig1}, the constitutive scalars $\gamma^{x,y,z}$
are plotted versus $z \in \le -t, t  \ri$ for $t=  \pi/6$ (green dashed curves),
 $\pi/3$ (red solid curves), and $4 \pi /9$ (blue broken dashed curves).
 The region $ \le 4 \pi /9 \ri  < t \leq \le \pi/2 \ri $, which includes the nonsingular Killing--Cauchy horizon solution at $t = \pi/2$, is avoided in Fig.~\ref{fig1} (and later  in Fig.~\ref{fig2}).
We have that $\gamma^x > 1$ while $\gamma^{y,z} < 1$ for  $t>0$. The constitutive scalars increasingly deviate from unity as $t$ increases from zero. In particular,  $\gamma^x$ very rapidly increases in magnitude as $t$ increases, while the magnitudes of $\gamma^y$
and $\gamma^z$ decrease very rapidly   as $t$ increases.
 Notice that $\gamma^y=\gamma^z$ in the plane $z=0$;
consequently, the Tamm medium is  uniaxial dielectric--magnetic in that plane.

\begin{figure}[!ht]
\centering
\includegraphics[width=3.5in]{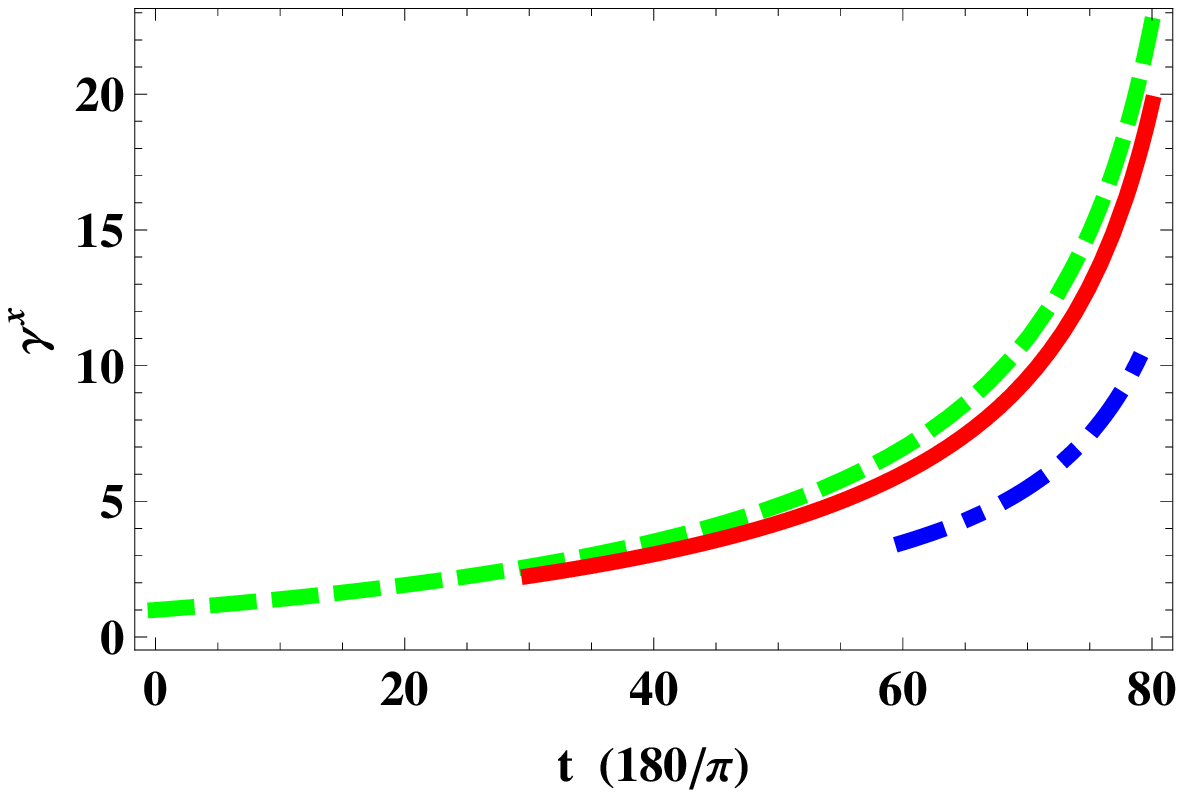}
\includegraphics[width=3.5in]{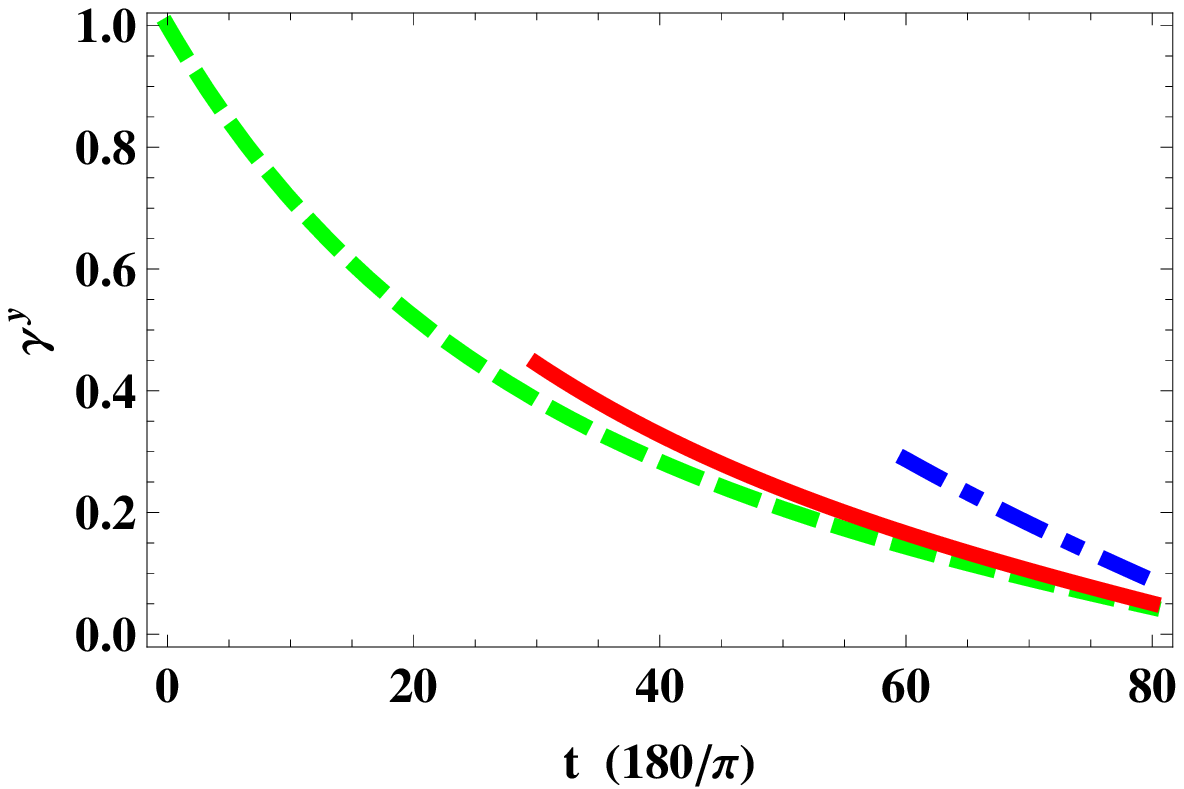}
\includegraphics[width=3.5in]{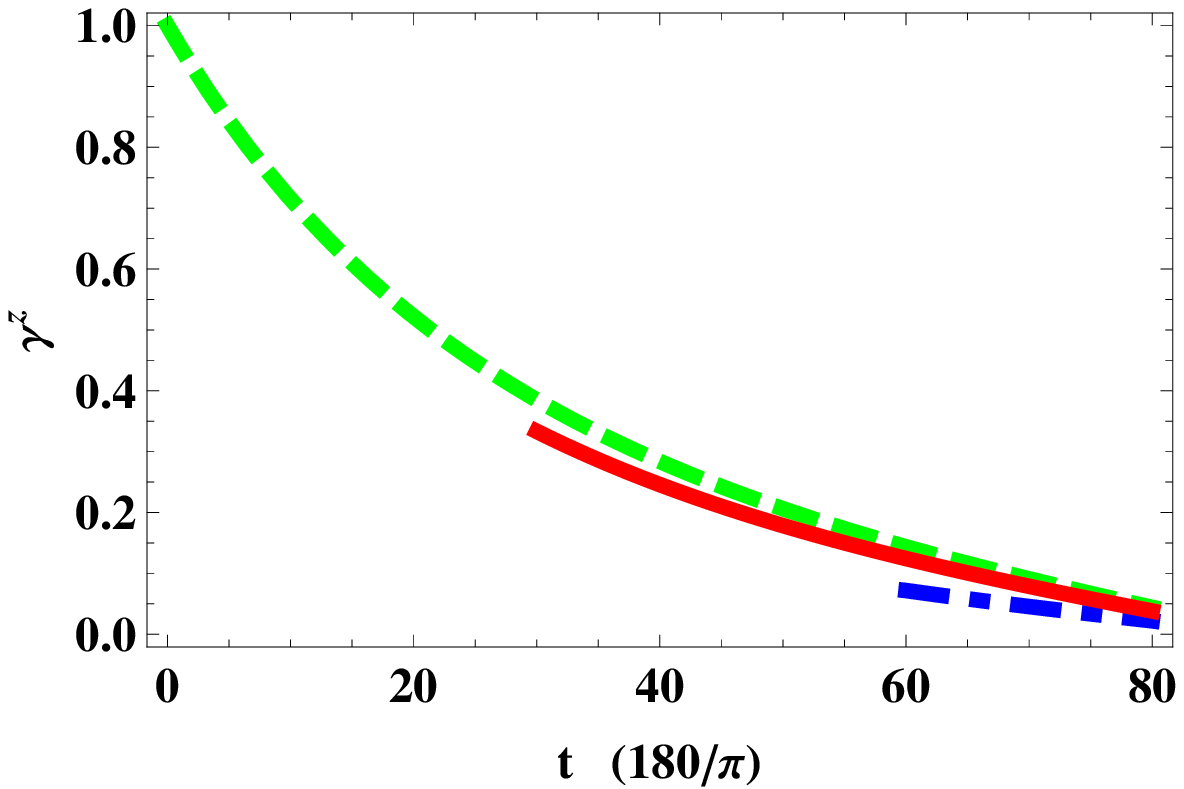}
 \caption{The parameters $\gamma^{x,y,z}$ plotted versus $t \in \le z, 4 \pi/9 \ri$ for $z= 0$ (green dashed curves),
 $\pi/6$ (red solid curves), and $\pi/3$ (blue broken dashed curves).
  \l{fig2}}
\end{figure}

A different perspective on the dependency of the constitutive parameters upon $z$ and $t$ is provided in
 Fig.~\ref{fig2},
where the constitutive scalars $\gamma^{x,y,z}$  are plotted versus $t \in \le z, 4 \pi/9 \ri$ for $z= 0$ (green dashed curves),
 $\pi/6$ (red solid curves), and $\pi/3$ (blue broken dashed curves). We see that  $\gamma^x$ becomes unbounded as $t$ approaches $\pi/2$, whereas   both $\gamma^y$ and $\gamma^z$ vanish as $t$ approaches $\pi/2$, for all values of $z$.

\section{Inverse Bruggeman formalism} \l{iBr}

The Tamm medium characterized by the constitutive parameters \r{Tamm_e_m} is an orthorhombic
 dielectric-magnetic medium with identical relative
permittivity and relative permeability dyadics. Such a medium
may be conceptualized as a homogenized composite material (HCM), using
 the well-established Bruggeman homogenization formalism \c{WLM_MOTL}.

We consider the homogenization of two component mediums,
identified by the labels $a$ and $b$.
Both  component mediums are isotropic
dielectric--magnetic mediums with relative permittivities $\eps_a$ and
$\eps_b$, and relative permeabilities $\mu_a$ and $\mu_b$. \blue{ In consonance with the relative permittivity and relative permeability dyadics of the Tamm medium being identical, we consider the component mediums characterized by  $\eps_a = \mu_a=\gamma_a$ and $\eps_b = \mu_b=\gamma_b$. Component mediums with such characteristics do not generally occur in nature but the sought after component mediums  could be envisaged as HCMs themselves, arising from the homogenization of isotropic dielectric and isotropic magnetic mediums \c{ML_PRB}.}

Both component mediums are
 distributed randomly, and their respective volume fractions are $f_a=1-f_b$ and $f_b \in \le 0,1 \ri$. Each component medium is
present as an assembly of electrically small ellipsoidal particles.
For example, at optical wavelengths, component particles with linear dimensions less than approximately 40--70 nanometers
would be required.
In keeping with the orthorhombic symmetry of the Tamm medium,
the major and minor axes of the component ellipsoidal particles are  aligned with the coordinate axes.
All ellipsoidal particles are assumed to have the same shape. Thus,
the vector
$\#r_{\,s} = \rho_\ell \, \=U  \cdot \hat{\#r}$, $\le \ell =  a, b\ri$,
prescribes
the surface of each ellipsoidal particle relative to its centroid, with
$\hat{\#r} = \le \sin \theta \cos \phi, \sin \theta  \sin \phi, \cos \phi \ri$ as the radial unit vector, and $\theta\in[0,\pi]$ and $\phi\in[0,2\pi)$ being the polar and azimuthal angles, respectively. A linear measure of the  size of the component ellipsoidal particles is provided by  $\rho_\ell$.  The shape of
the component  particles is captured by
  the positive-definite  dyadic
$\=U  =
 \mbox{diag} \le U^x , U^y, U^z  \ri$.

The Bruggeman formalism requires
 the polarizability density dyadics \c{WLM_MOTL}
\begin{equation}
 \=a_{\,\ell} = \le \gamma_\ell \=I - \=\gamma \ri
\cdot \les \, \=I + \=D_{\,\ell} \cdot \le \gamma_\ell \=I -
\=\gamma \ri \ris^{-1} ,\qquad  \le \ell = a, b  \ri.
\end{equation}
  The depolarization dyadics $\=D_{\, \ell}$ herein
  are given by the  double integrals \c{Michel,Michel_WSW}
  \begin{eqnarray} \l{D_dint}
&&\=D_{\, \ell} = \frac{1}{4 \pi} \int^{2 \pi}_\phi d\phi \int^\pi_\theta d \theta \, \sin \theta\, \frac{ \le \=U^{-1}  \cdot \hat{\#r} \ri \le \=U^{-1}  \cdot \hat{\#r} \ri}{\le \=U^{-1}  \cdot \hat{\#r} \ri \cdot \=\gamma  \cdot \le \=U^{-1}  \cdot \hat{\#r} \ri} , \nonumber \\ && \hspace{60mm} \le \ell = a, b  \ri\,,
  \end{eqnarray}
which may be conveniently represented in terms of incomplete elliptic integrals \c{WSW}.
The constitutive parameters of
the HCM
 are related to the constitutive  and shape parameters, and volume fractions,  of the component mediums by the  Bruggeman
equation \c{EAB,WLM_MOTL}
\begin{equation}
\label{Brug}
 f_a \,\=a_{\,a} + f_b\, \=a_{\,b}  =\=0\,,
\end{equation}
which is equivalent to three independent  scalar equations, which are
coupled via the constitutive parameters for the HCM and the shape parameters for the component mediums.

Without loss of generality, we fix $U^x = 1$. Then we recast Eq.~\r{Brug} to solve the following problem: With $\=\gamma(z,t)$ known,
determine $\gamma_a(z,t)$, $f_a(z,t)$ and $U^y(z,t)$ in terms of $\gamma_b(z,t)$  and $U^z(z,t)$ for any combination of $z$ and $t$ in Region 1.
The inverse Bruggeman
formalism
 is then conveniently implemented for any specified $z$ and $t$  by
exploiting direct numerical methods  \c{ML_JNP}.

\section{Numerical results}\label{nr}

Let us now provide representative numerical illustrations of
component mediums which could be homogenized to realize the Tamm medium specified by Eqs.~\r{Tamm} and \r{Tamm_e_m},
based on the inverse Bruggeman formalism  described in Sec.~\ref{iBr}. The Tamm medium for the Region I spacetime was considered for two cases: the first concerns fixed $t$  while the second concerns fixed $z$. For both cases the immediate vicinity of the Killing--Cauchy horizon at $t= \pi/2$ was excluded, in order to avoid  extreme constitutive parameter regimes  for the Tamm medium.

\begin{figure}[!ht]
\centering
\includegraphics[width=3.5in]{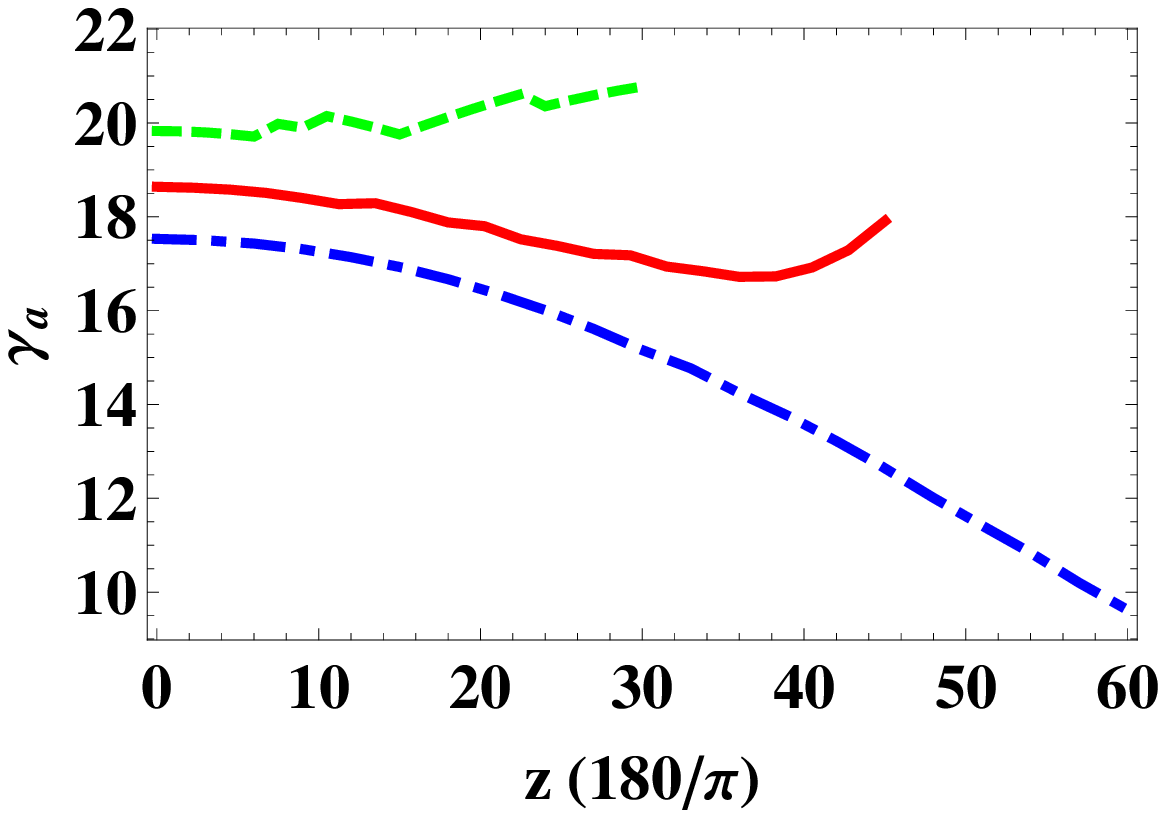}
\includegraphics[width=3.5in]{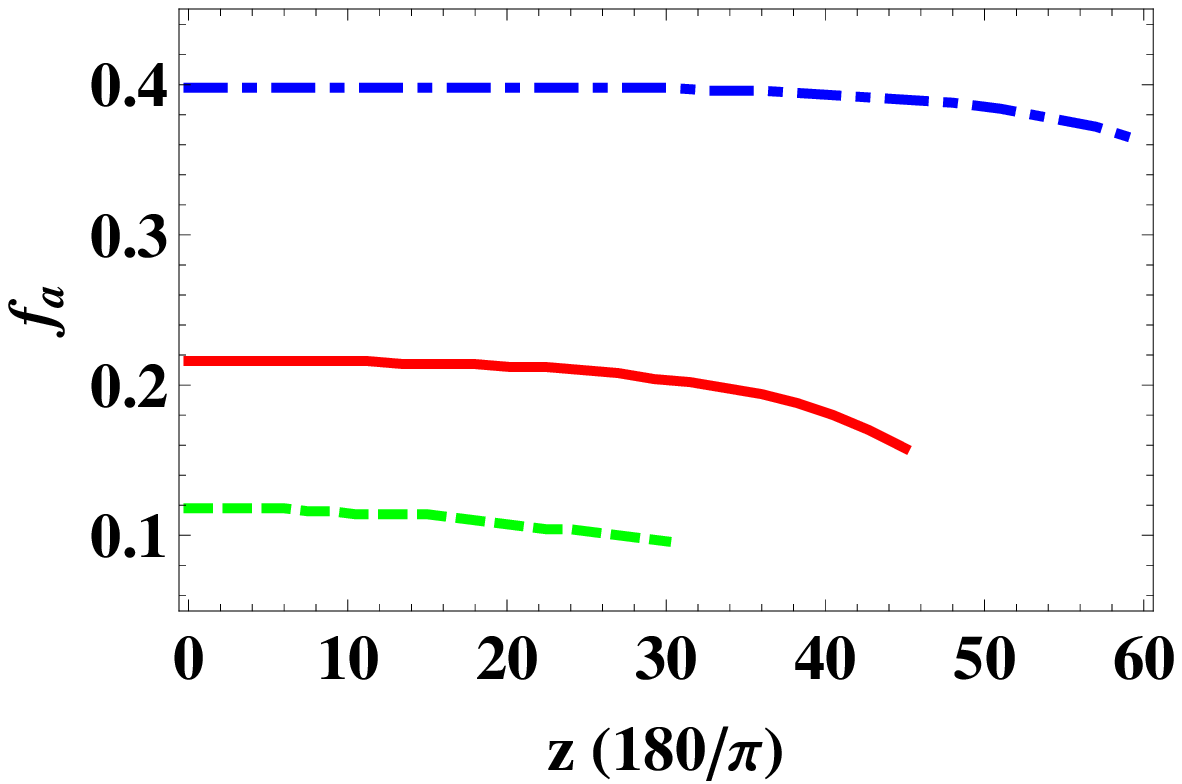}
\includegraphics[width=3.5in]{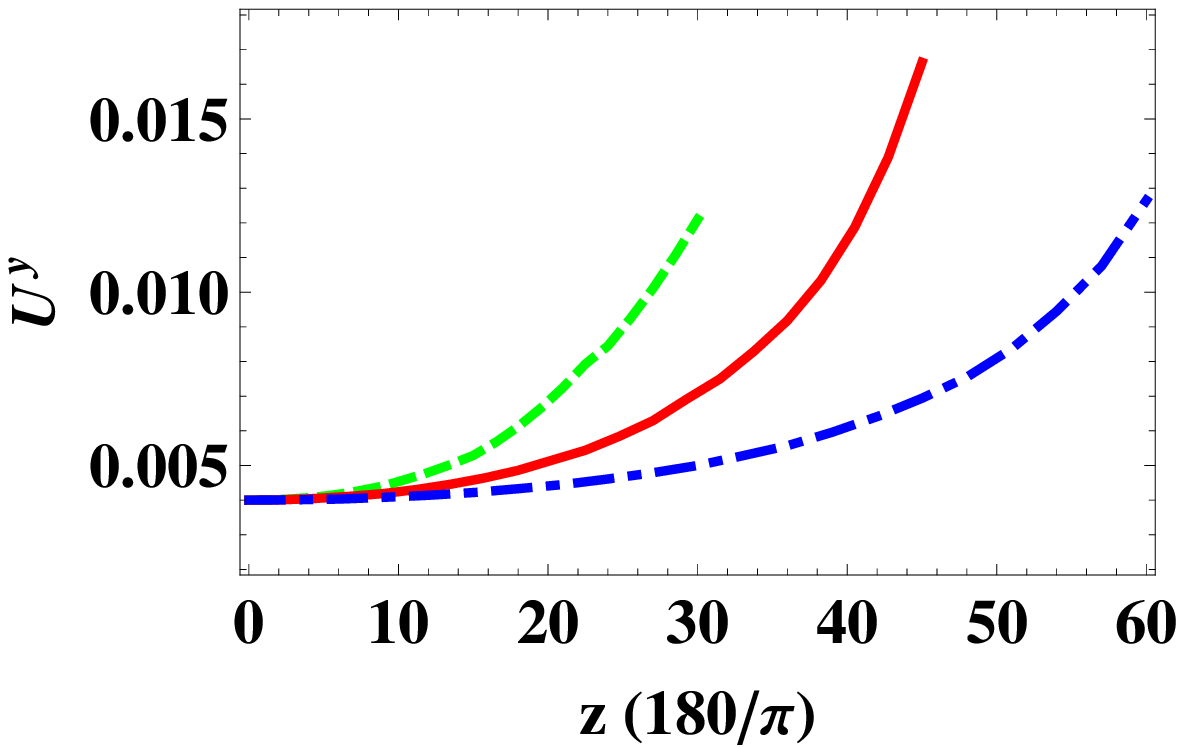}
 \caption{ The parameters $\gamma_a$,   $f_a$, and   $U^y$
  plotted versus $z \in \le -t, t  \ri$ for $t=  \pi/6$ (green dashed curves),
 $\pi/4$ (red solid curves), and $\pi /3$ (blue broken dashed curves).
  \l{fig3}}
\end{figure}

 We
 begin by considering the Tamm medium at fixed $t$. Estimates of  $\gamma_a$, $f_a$ and $U^y$ are plotted against $z \in \le 0, t \ri$ for $t = \pi/6$, $\pi/4$ and $\pi/3$ in Fig.~\ref{fig3}. The shape parameter $U^z  = 0.004$ for these calculations, while $\gamma_b  = 0.3$ for $t= \pi/6$, $\gamma_b  =0.14 $ for $t= \pi/4$, and $\gamma_b  = 0.03 $ for $t= \pi/3$.
  The values of
$U^y$ rise sharply as $z$ increases, for all values of $t$ considered.
 In the limit $z \to 0$,  for all values of $t$ the values of the shape parameter $U^y$  converge on $0.004$, which is the value of $U^z$. This is a
reflection of the fact that the Tamm medium is a uniaxial dielectric--magnetic medium in the $z=0$ plane. The variation in    $\gamma_a$ as $z$  varies
depends upon the value of $t$: at larger values of $t$, $\gamma_a$ is more sensitive to changes in $z$. The volume fraction $f_a$ varies  relatively little as  $z$ varies.

\begin{figure}[!ht]
\centering
\includegraphics[width=3.5in]{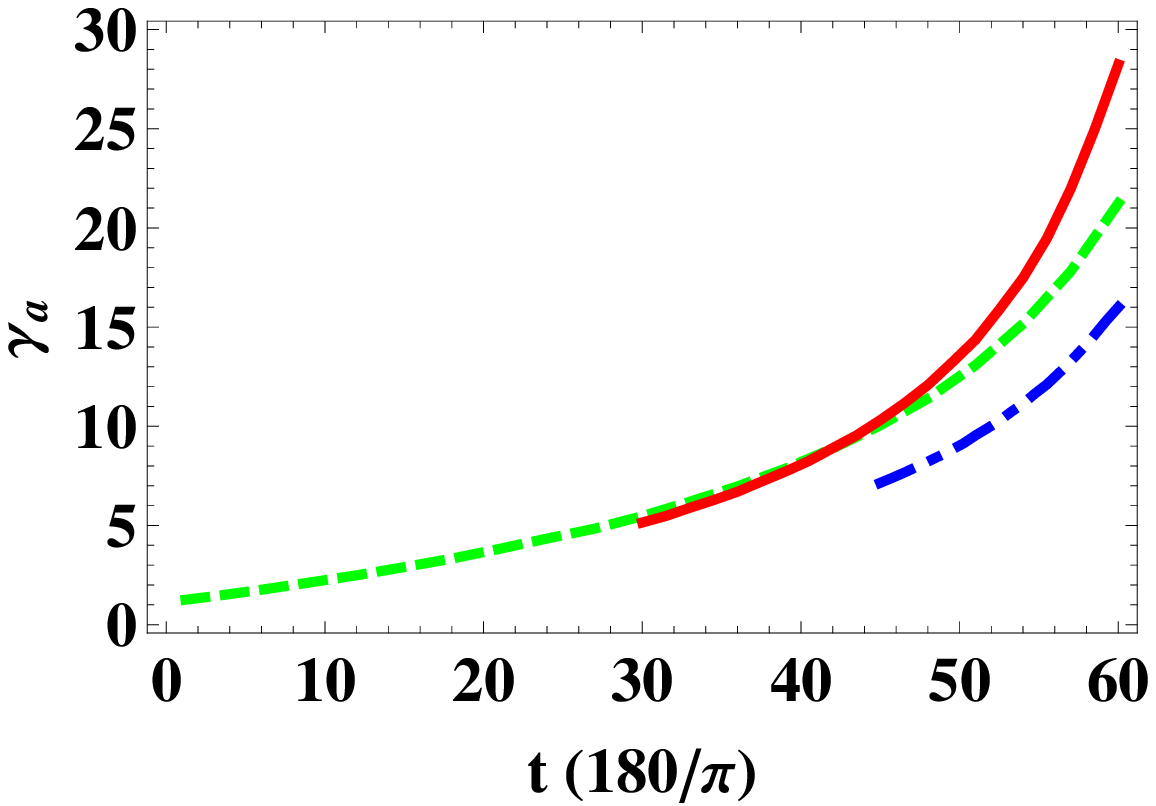}
\includegraphics[width=3.5in]{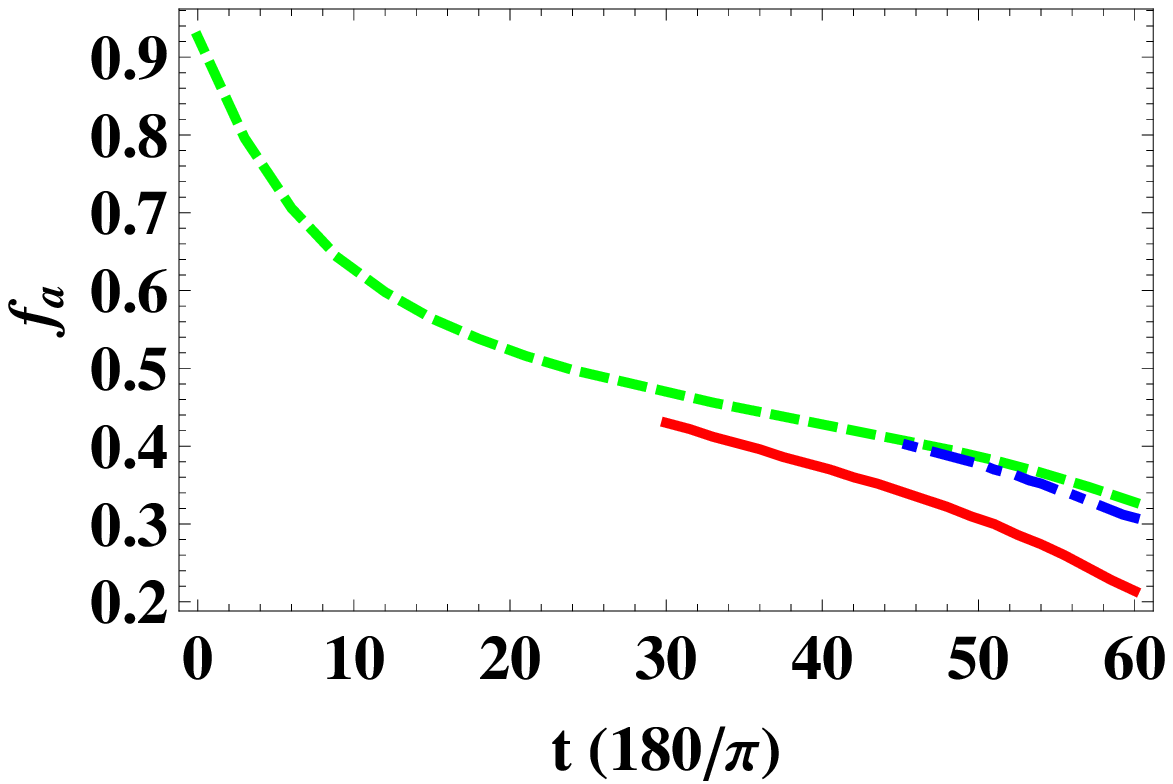}
\includegraphics[width=3.5in]{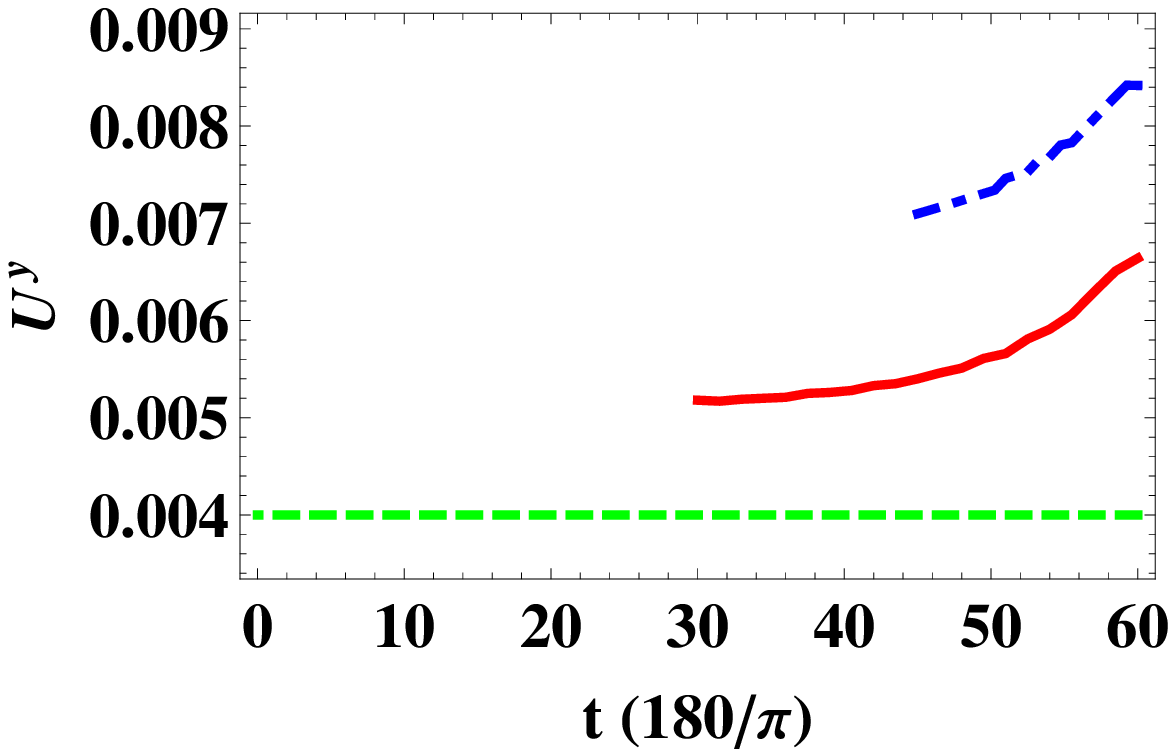}
 \caption{The parameters $\gamma_a$,   $f_a$, and   $U^y$
  plotted versus $t \in \le z, \pi/3  \ri$ for $z=  0$ (green dashed curves),
 $\pi/6$ (red solid curves), and $\pi /4$ (blue broken dashed curves).
  \l{fig4}}
\end{figure}

Next, let us consider the Tamm medium at fixed $z$. In Fig.~\ref{fig4}
estimates of  $\gamma_a$, $f_a$ and $U^y$ are plotted against $t \in \le z, \pi/3 \ri$ for $z = 0$, $\pi/6$ and $\pi/4$.
 Here  $\gamma_b  = 0.05$ for $z= 0$, $\gamma_b  =0.08 $ for $z= \pi/6$, and $\gamma_b  = 0.05 $ for $t= \pi/4$.
 As in Fig.~\ref{fig3}, the shape parameter $U^z = 0.004$. When $z=0$, the shape parameter $U^y$ again remains constant at $U^z$. Aside from this exception, the parameters $\gamma_a$, $f_a$ and $U^y$ all vary markedly as $t$ increases, for all values of $z$ considered.

\section{Discussion}\label{disc}

Graded dispersals of electrically small ellipsoids of two isotropic dielectric--magnetic component mediums
may be used to realize the Tamm medium representing
the collision of two  linearly polarized gravitational
plane waves. By varying
the volume  fraction and the relative permittivity (or
 permeability) of each component mediums, along with
the common ellipsoidal shape,
Region I in spacetime can be accessed except the immediate vicinity of the Killing--Cauchy horizon at $t= \pi/2$.
Thus, the inverse
homogenization formulation paves the way for
 an experimental analog for the collision of two  linearly polarized gravitational
plane waves. For example, the experimental analog could be used to study either the temporal evolution of the colliding plane waves at a fixed location, or the spatial portrait of the colliding plane waves at  a fixed time. \blue{ The nonhomogeneous nature of the underlying spacetime may be accommodated by stitching together adjacent spacetime regions that are sufficiently small that they could be approximated by homogeneous Tamm mediums.
Experimental realization for optical experimentation would then involve nanocomposite materials \cite{M_JNP_extended}.}

\blue{
The presented metamaterial model is essentially wavelength independent. Frequencies higher or lower than optical frequencies could be used in order to access regimes where the  values of relative permittivity and/or relative permeability required for the component mediums may be more readily attained. However, if higher frequencies were used, then the component medium particles would have to be smaller in order for the concept of homogenization to remain valid \c{EAB}.}

The particular  inverse homogenization formulation chosen for presentation in
Sec.~\ref{iBr} was based on two isotropic dielectric--magnetic component mediums, each being an assembly of oriented ellipsoidal particles. However,
alternative formulations may be envisaged. For example,
the homogenization of four component mediums could also yield an HCM representing the Tamm medium. In this case, two isotropic dielectric mediums and two isotropic magnetic mediums, with each component medium comprising oriented spheroidal particles,
could be used; or instead two uniaxial dielectric mediums and two uniaxial magnetic mediums, with each component medium comprising spherical particles,
could be used \c{TGM_WSW}.

Let us comment on the restriction of the  inverse
homogenization approach to
Region I in spacetime away from the immediate vicinity of the
 Killing--Cauchy horizon at $t= \pi/2$. \blue{ As the point $t= \pi/2$ is approached,
 the Tamm medium becomes extremely anisotropic (as may be appreciated from Figs.~\ref{fig1} and \ref{fig2}). Accordingly, component mediums specified by  extreme values  of relative permittivity (and permeability)---i.e., both extremely high values and values exceedingly close to zero---would be needed to implement the corresponding HCM. At first glance,
the realization of such parameter values would pose a major challenge to experimentalists.
However, the relative permittivities
(and  permeabilities) represented in Figs.~\ref{fig3}--\ref{fig4}
are not beyond the grasp of contemporary materials researchers, and these correspond to a wide area of the Region I spacetime.
 Indeed,  considerable progress been made within recent years as regards metamaterials with
extremely high relative permittivities and permeabilities \c{High_n} and with
relative
permittivities and relative permeabilities very close to zero \c{Cia_PRB}.  Furthermore, extremely high degrees of electric and magnetic anisotropy can be achieved via the homogenization of highly elongated component particles \c{Gevorgyan,Mackay_arxiv}.}

\blue{
Lastly,  in principle, the metamaterial model described here could be extended, using the same techniques as presented in Secs.~\ref{Tamm-medium} and \ref{iBr},  to  rather more complicated scenarios; i.e., ones even less well suited to analytical and numerical study than the collision of gravitational plane waves described herein. For example, by combining with  results presented previously \c{ML_PRB},
 the collision of gravitational plane waves in a Schwarzschild-(anti-)de Sitter spacetime background could be contemplated.}

%
%

\end{document}